\documentstyle[prd,preprint,aps]{revtex}

\newcommand{\be}{\begin{equation}}
\newcommand{\ee}{\end{equation}}
\newcommand{\bea}{\begin{eqnarray}}
\newcommand{\eea}{\end{eqnarray}}

\begin{document}

\reversemarginpar
\tighten

\title{A not so brief commentary \\ on \\ cosmological entropy bounds}

\author {A.J.M. Medved}

\address{
School of Mathematics, Statistics and Computer Science\\
Victoria University of Wellington\\
PO Box 600, Wellington, New Zealand \\
E-Mail: joey.medved@mcs.vuw.ac.nz}

\maketitle

\begin{abstract}

There has been, quite recently, a discussion on how holographic-inspired bounds
might be used to encompass 
the present-day dark energy and early-universe inflation
 into a single paradigm.
In the current treatment, we point out an inconsistency in  
the proposed framework
and then provide a viable resolution. We also elaborate on
 some of the implications of this
framework and further motivate the proposed holographic connection. 
The manuscript ends
 with a more speculative note on cosmic time as an emergent
(holographically induced) construct.

\end{abstract}

\section{Background and Motivation}

The holographic principle   can
reasonably  be viewed as a semi-classical manifestation
of quantum gravity that places a fundamental limit
on informational-storage capacities \cite{Bou-1}.
To translate this notion of holography into a cosmological setting
has been an ongoing concern of the theoretical community;
beginning with the pioneering work of Fischler and Susskind
\cite{FS} (also see, {\it e.g.},
  \cite{Ven,Hol-cos2,Hol-cos3,Hol-cos4,Bru-2}).
For the most part, the challenge has been  to find  
a way of placing a meaningful and well-defined upper bound 
on the  entropy ({\it i.e.}, the potential for information) in
various cosmological situations.
 Optimistically speaking, one would prefer a  bound that limits the entropy
on  a {\it spacelike} hypersurface; in particular, 
the  observable universe at a specified moment in time.
Certainly, this issue has been addressed by
 many interesting strategies and proposals.
Nonetheless, progress in this direction has been impeded by the ambiguous  
nature of both entropy
and the holographic principle itself. 

An especially pertinent question is what (if anything) holography can 
tell us  about inflationary~\cite{Inf} fluctuations. This could
be of interest at early times in the era, 
when  inflation  can be modeled as an effective (quantum) field theory within
the confines of the horizon; as well as
at later times, after many of the fluctuations
have been classically frozen at super-horizon scales. As a for instance,
it has been asked if holographic bounds can enable
us to place an upper limit on the ultraviolet-cutoff scale of the effective
field theory
(for relevant work, see \cite{Hol-inf1,FK,Dan,KVS,Hol-inf5}).
One might also inquire as to  the viability of holographically implicating 
 the so-called dark energy~\cite{DE}; that is, the exotic
energy source that mimics  a cosmological constant~\footnote{It is,
of course, within the realm of possibility that the dark energy 
mimics a cosmological 
constant because
it is a cosmological constant.}
and  is responsible for our current period of acceleration.
In fact, it has become fashionable, as of late, to call upon  
holographic arguments for the purpose of 
rationalizing  the (inexplicably) small magnitude of the dark energy 
({\it e.g.}, \cite{CKN,Hol-de2,Hsu,Li,Hol-de5}).

Let us emphasize that  early-universe inflation also represents
a cosmic period of acceleration; hence, there appears to be some sort of   
relationship ---  if only  a  superficial one --- between
the inflationary potential and  the dark energy.
It is therefore quite natural, in the context of our discussion,
to ask if this notion of a  connection could be strengthened by appealing
to the holographic paradigm.  In this regard, let us point out
a recent paper of interest by Myung  \cite{Myu-1}, who 
discussed how  a particular holographic framework
may  be used to encompass both of these cosmic events.
To elaborate,~\footnote{In the  first two sections,
 our attention will mainly be directed at
the {\it framework} used in \cite{Myu-1}. Meanwhile, we will defer
 discussing  the
(proposed) holographic connection, itself,   
until Section III.} 
this author primarily focused on  a pair of holographic
bounds; with these being  originally proposed by E. Verlinde
for a closed, radiation-dominated (Friedmann-Robertson-Walker) universe 
\cite{Ver},
and later  generalized by Youm  to incorporate 
flat and open topologies as well \cite{You-1}.
Guided by these antecedent treatments, the author \cite{Myu-1} 
begins by
distinguishing between a {\it weakly} gravitating
 and a {\it strongly} gravitating phase of 
the universe.
Essentially, $HR<1$ corresponds to the former
regime and  $HR>1$, the latter; with 
$R$ being the cosmological scale factor,
$H= {\dot R} /R$ being the Hubble parameter (assumed to be positive),
and a  dot representing a differentiation with respect to cosmic 
time. [Technically speaking, the product $HR$
should be compared with $\sqrt{2-k}$, where $k$ takes
on a value of $-1$, $0$ and  $+1$ for an open, flat
and closed universe respectively \cite{You-1}.
We will, however, consistently ignore numerical
factors  of the order unity throughout the
paper.  Some further remarks on current conventions: We have chosen
to work with a four-dimensional universe, although 
anything said in this paper can  readily be extended
to higher-dimensional models. Also, all fundamental
constants --- except for  the Planck mass,
$M_p=G^{-1/2}$ --- will always be set to unity.]

After discriminating  between weak and strong gravity, 
one may then utilize,
as prescribed by  \cite{Ver,You-1}, a
respective pair of  holographic bounds:
\be
S <  S_{B} \sim RE \quad\quad\rm{if}\quad HR<1
\label{1}
\ee
and
\be
S < S_{H} \sim {VH\over G} \quad\quad\rm{if}\quad HR>1\;,
\label{2}
\ee
where $S$ and $E$ are the universal entropy and
energy respectively, and $V\sim R^3$ is the volume. [Actually, Myung prefers
to replace  equation (\ref{1}) with the following bound on the energy:
$E<R/G$, so that  $E$ is limited by the Schwarzschild radius
of a universal-sized black hole. 
But consider that, in the case of ``very weak''
gravity  or $HR<<1$, it is commonly recognized that $S<S_{B}$ is 
the  strongest 
(viable) holographic constraint \cite{Bek-2,Bou-1};
while near the saturation point or $HR\sim 1$ (which is really the main  
focus of either \cite{Myu-1}  or our subsequent discussion), 
this energy bound  and equation (\ref{1})
 are 
parametrically equivalent \cite{Ver}.  Given these observations,
we can see no good reason for dispensing with the usual convention
of expressing holographic bounds in terms of the entropy.]

Before we proceed onwards,
some elaboration on these bounds would appear to be in order.
First of all, $S_B$ essentially represents  
Bekenstein's famed entropy bound \cite{Bek-2}, which
is believed by many to be a  necessary restriction on any 
weakly gravitating system ({\it e.g.}, \cite{Bou-2}).~\footnote{The Bekenstein
 bound is not,  however,  without its
fair share of controversy; {\it e.g.}, \cite{Marolf,Page}.}  
Now let us consider near
 the saturation point,
where the system should just be on the cusp of  black hole formation
($E\sim R/G$).
Here, it becomes evident that $S_B\rightarrow S_{BH}\sim R^2/G$ as
$HR\rightarrow 1$;
where $S_{BH}$ is readily identifiable
as the Bekenstein--Hawking entropy  
\cite{Bek,Haw} of a universal-sized black hole.
It follows that
$S< S_{BH}$ must be true for any weakly gravitating system,
which is obviously an anticipated result. 
Meanwhile, $S_{H}$ can be interpreted  as 
the amount of entropy that is stored in a volume
of space $V$ when it has been filled-up with  Hubble-sized black holes
(since  $V/H^{-3}\;\times\; H^{-2}/G \sim S_H$).
Common-sense considerations suggest that 
this would likely be the largest allowable  entropy in a strongly
gravitating region, inasmuch as $H^{-1}$ is expected
to be the maximally sized {\it stable} black hole under 
such conditions. In fact, the so-called
Hubble bound of equation  (\ref{2}) has, like the Bekenstein bound,
 been proposed in literature 
\cite{Ven} that
predates Verlinde's treatment.

\section{Critique and Resolve}

To reiterate a pivotal point, the Verlinde--Youm 
holographic bounds  were central to the 
discussion in \cite{Myu-1} on
inflationary perturbations, current-day dark energy and their
possible connection. Although 
sympathetic to the general philosophy of \cite{Myu-1},
we feel there are  two critical issues that still need to 
be addressed. Firstly, the Verlinde--Youm framework (and, hence, any
resultant bound) is specific
to a {\it radiation-dominated} universe.  As explicitly shown in \cite{You-2},
for a universe with an arbitrary equation of state 
($p=\omega \rho$),~\footnote{As per usual conventions, we assume perfect-fluid
matter for which  $p$ is the pressure,
$\rho$ is the energy density, and $\omega$ is the equation-of-state parameter.
(For future reference, $T$ will be the temperature.)
Note that $\omega$ equals $+1/3$ for pure radiation,
$0$ for dust, must be less than $-1/3$ for
an accelerating universe, and saturates  $-1$ for a true
 cosmological constant. The present-day observed value happens to be
bounded by $\omega<-0.75$ \cite{DE}.}
the analogous bounds take on much different form 
when  $\omega$  differs from its radiative value
of $+1/3$. It should be clear that, for the  the eras of particular interest 
(early inflation and the current acceleration),
the universe cannot possibly be regarded  as being  radiation
dominated. Secondly and much more aesthetically, it can perhaps be argued that
a truly encompassing consequence of  holography   should be reducible
to a single entropic bound. In this sense, we find the
distinction between strongly and weakly gravitating regimes to be
somewhat unsettling.

Actually, one can resolve the second issue (but not the first) by
returning to the seminal paper by Verlinde  \cite{Ver}.
There, it was  argued that  the bound $S_C<S_{BH}$  --- with $S_C$  
representing an appropriately defined   {\it Casimir} entropy ---
is universal in the sense of being equally applicable to regimes 
of strong and weak gravity. Which is to say, the Casimir entropy
of the universe
should {\it never} be able to exceed that of a universal-sized black hole.
 Let us  be a bit more explicit about what is meant, in this particular
context, by the Casimir entropy.  
To arrive at this quantity, one  first  identifies a
Casimir energy $E_C$
with   the violation of the thermodynamic Euler identity (leading
to $E_C\sim k[E+pV-TS]\;$ \cite{Ver,You-1}), 
and then relates
$S_C$ to $E_C$ by way of the {\it saturated} Bekenstein bound 
({\it i.e.}, by fixing $S_C\equiv R E_C $ up to a  neglected
 numerical factor). 
As might have been  anticipated,
the Casimir entropy then  turns out to be  
a sub-extensive quantity \cite{Ver}, 
scaling rather like an area! It is, therefore, not difficult
to see how the bound $S_C<S_{BH}$ could very well have  universal
applicability, at least for the case of a closed universe.
 Unfortunately,
this definition does become problematic
for  a flat or open universe, insofar as $E_C\propto k$;
meaning that $S_C$ would then be vanishing or negative (respectively).
On the other hand, one only needs to follow Youm's prescription \cite{You-1}
to obtain a Casimir entropy that is strictly positive
for all relevant topologies. This generalization does, however,
seem to be somewhat contrived. 

As  alluded to above, Verlinde's bound on the Casimir entropy resolves
 our second point of contention but not our first. 
However, we can proverbially
kill two birds with one stone by calling upon a closely
related but yet distinct entropy bound.  Here, we have in mind 
a bound that was  originally
proposed by Brustein {\it et al} \cite{Bru-1} and  can essentially be viewed as
an incarnation  of the (so-called) causal entropy bound \cite{Bru-2}.
The causal entropy bound can, in turn, be  
accurately perceived as a generalization of 
the Hubble bound. (More specifically, to obtain the causal 
bound from $S_H$, one  replaces the heuristic  length scale $H^{-1}$ with
a covariantly and rigorously defined ``causal-connection scale''.)
To elaborate on the logistics, let us  follow \cite{Bru-1} and
first define  a {\it sub-extensive} entropy ($S_{SUB}$)
in accordance with the relation
\be
S=\sqrt{2S_BS_{SUB}}\; ;
\label{4}
\ee
which is to say, $S_{SUB}\sim S^2/ER$.
Then, as documented in \cite{Bru-1},
the causal entropy bound can  be used to verify the validity ---
under quite generic circumstances --- of the following holographic bound:
\be
S_{SUB} < S_{BH}\sim {V\over G R}\quad\quad {\rm for\;any\;value\;of}\;\; HR 
\;.
\label{5}
\ee

There are a few points of interest regarding this ``improved'' form of
sub-extensive entropy  bound; namely: \\

{\it (i)} Comparing equation (\ref{4}) with the Cardy--Verlinde formula
or \cite{Ver} 
\be
S=\sqrt{2S_BS_C-S_C^2}
\label{6}
\ee
(as applicable to a closed, radiation-dominated universe),
one finds that  $S_{SUB}$ and $S_{C}$ are in agreement   up
to terms of the order $S_C/S_B$. Nevertheless, $S_{SUB}$ must be viewed
as the preferable choice on which to base an entropy bound,
 being a quantity that maintains its integrity
irrespective of the equation of state, topology,
{\it etcetera}. \\

{\it (ii)} It is also prudent to compare equation (\ref{4})
with  Youm's generalization of the Cardy--Verlinde formula or \cite{You-1} 
\be
S=\sqrt{2S_BS_C-kS_C^2}
\label{7}
\ee
(as  applicable  to a radiation-dominated universe that is closed,
open or flat, and recall that  $S_C$ is defined so as to be a 
strictly positive quantity). 
Interestingly,
 $S_{SUB}$ and $S_C$ happen to  agree perfectly for the $k=0$ case
of a flat universe. Yet, any bound that is
 based on  $S_C$ remains problematic,
insofar as neither equation (\ref{6}) nor
(\ref{7}) holds up when $\omega$ differs from $+1/3$ \cite{You-2}. 
That is,  one cannot  justifiably expect, {\it a priori},
that  the Verlinde--Youm formalism will directly  translate into a
context of cosmological  acceleration. \\

{\it (iii)} It can be show that, for a regime of weak gravity ($HR<1$),
the ``improved'' bound of equation (\ref{5}) ensures
the validity of the Bekenstein bound (\ref{1}).
To see that this is correct, let us begin by employing
 $S_{SUB}\sim S^2/ER$  [{\it cf}, equation (\ref{4})]
to  rewrite equation (\ref{5}) as follows:
\be
S< \sqrt{VE\over G}\;,
\label{8}
\ee
which is --- not coincidentally ---  yet another incarnation of the
causal entropy bound \cite{Bru-2}. Now using the weak-gravity constraint,
as well as the first Friedmann equation or 
$H^2\sim G E/V$,~\footnote{Although we work with a flat universe for 
simplicity,
 the results of both (iii) and (iv) readily carry through for an open
or closed universe, with $H^2+kR^{-2}\sim GE/V$. 
One can confirm this assertion
 by inserting all the numerical factors and then
 imposing the ``technically
correct'' definition of weak (strong) gravity; that is,  $HR<\sqrt{2-k}$
$(HR>\sqrt{2-k}$) \cite{You-1}.}
we have
\be
{\sqrt V} > R\sqrt{GE}\;.
\label{9}
\ee
 Combining these last two equations, we immediately
obtain the Bekenstein bound (\ref{1}). \\

{\it (iv)} Moreover, the improved bound of equation (\ref{5})
guarantees the validity of the Hubble bound (\ref{2}) when
the system is strongly gravitating.  For the purpose of verifying this 
outcome, 
let us again utilize the first  Friedmann equation, along
with the bound $HR>1$, to obtain
\be
{\sqrt E} > R^{-1}\sqrt{V\over G}\;.
\label{10}
\ee
Substituting this result into equation (\ref{8}), we then find
\be
S<{V\over GR}\;.
\label{11}
\ee
But, since $R^{-1}<H$ for a strongly gravitating system, 
the Hubble bound (\ref{2}) necessarily follows. \\

It is these last two points, in particular, that provide the
(previously missing) justification for one to
 apply the holographic
bounds (\ref{1},\ref{2}) in an inflationary or dark-energy context. 
Which is to say, these equations can now be motivated by the
 ``improved'' holographic bound of equation
(\ref{5}), which  has been argued to have universal validity \cite{Bru-1}.
Conversely, the same claim can {\it not} be made, in good faith,
about the Verlinde--Youm framework, which cannot necessarily
be extrapolated away from radiation-dominated systems.~\footnote{On this
note, we should point out that Myung \cite{Myu-1} did briefly discuss another 
Verlinde-inspired 
 holographic bound. This being  $T>-{\dot H}/(2\pi H)$  when $HR>1$ 
\cite{Ver}. 
 Unfortunately, we are yet unable to substantiate this bound by appealing
to equation (\ref{5}). Nevertheless, this lower temperature bound
is actually just a  restatement of one of  the slow-roll conditions
of inflation \cite{Myu-2}, so it almost certainly has validity in (at least) 
an inflationary context.} Moreover, by reducing the framework to
just the single entropic bound (\ref{5}), we are able to punctuate
the  argument \cite{Myu-1} that
temporally well-separated  systems can have a close holographic connection.

\section{Implications and Speculations}

For the sake of closure, let us now  elaborate on 
some of the holographic implications in the  cosmological
regimes of interest. We will consider, in turn,
the field-theoretical (or sub-horizon) inflationary perturbations, 
the classically frozen (or super-horizon) inflationary perturbations, and
the present-day dark energy.
(For further, partially overlapping discussion,
see \cite{Myu-1}.~\footnote{Also, for a particularly rigorous treatment
on holography and the dark energy, see \cite{BNOV}.})
We end with some summarizing and speculative comments. 

\subsection{Sub-Horizon Perturbations}

Here, we would like to begin by  assigning an entropy to the inflationary
(perturbative)
modes at sub-horizon scales or, equivalently, at early times in the
era. For this purpose, one normally proceeds ---
in lieu of a definitive explanation for the ``inflaton'' ---
by  adopting the pragmatic model of
an effective (quantum) field theory.  However, even within this
simplified framework,
 it is still not entirely clear as to how the entropy should be
quantified.  This ambiguity is a reflection of
the global picture; which is  essentially that of  a pure state
and, consequently,  one of vanishing entropy. 
It follows that any meaningful  calculation of the local (sub-horizon) entropy
will necessarily require  a suitable process of
 {\it coarse graining}.
(For  recent discussion on these and related matters, 
see \cite{Dan,KVS}.)

Conveniently for us, 
Gasperini and Giovannini \cite{GG} have already accomplished the stated task; 
 having  related the  entropy (per comoving mode)
to the ``squeezing'' that is induced by the time-dependent
(and, hence, particle-creating) background spacetime.  When all is said
and done, they were  able to express this coarse-grained  entropy
in terms of the Hubble parameter, as well as  an ultraviolet (energy) cutoff   
for the effective field theory.  More explicitly \cite{GG},
\be
S_{cg}\sim {\Lambda^2\over H^2}\;,
\label{12}
\ee 
where
  $\Lambda$ represents the ultraviolet cutoff and we have, as per usual,
ignored the numerical factors.

Now what about holography? As appropriate for a local observer during
the early stages of inflation,
we should restrict considerations to the region $R<H^{-1}$; meaning,
a regime of weak gravity.  Hence, this field-theoretic (coarse-grained) 
entropy is required  to satisfy the Bekenstein bound. 
So let us insert  equation
(\ref{12}) into the bound (\ref{1}) and then replace $E$ via the Schwarzschild
(weak-gravity) 
constraint $E<RM_p^2$; thus obtaining
\be
{\Lambda\over H}< M_p R\;.
\label{13}
\ee
It stands to reason that this bound should be  closest
 to saturation when $R\rightarrow H^{-1}$. To put it another way, 
$H^{-1}$ ---  being  the proper size of the  horizon ---   
can also  be viewed  as  the implicit
infrared (length) cutoff for the field theory. Hence, we can 
expectantly write
\be
\Lambda < M_p\;.
\label{14}
\ee
Insofar as the Planck mass is the  
natural energy scale for quantum gravity, this
restriction on $\Lambda$ is reassuring but, all by itself, not particularly
useful.

There is, however, an existent viewpoint that
the ultraviolet and infrared cutoffs should not, in spite
of appearances, be regarded as independent quantities. 
As first argued by Cohen {\it et al} \cite{CKN}, 
these cutoffs should adhere to
\be
\Lambda^4 L^3 \lesssim M_p^2 L\;.
\label{15}
\ee
This relation follows, quite simply, from the notion that
the maximal energy of the effective field theory (the left-hand side)
is not large enough to  collapse the region
into a black hole. Hence, one obtains the (saturating) relation 
$\Lambda^2\sim M_{p}/L$; meaning that, for $L\sim H^{-1}$,
\be
\Lambda\sim\sqrt{M_p H}\;.
\label{16}
\ee
The essential point here is that this outcome can also be rationalized
in the context of our current framework:  Firstly, equation (\ref{14})
tells us that a field-theoretic description is indeed justifiable
throughout the entire  inflationary phase. Hence, $E\sim\Lambda^4 R^3$
should provide an accurate description of
the energetics, and so the Bekenstein bound (\ref{1})
can just as legitimately be expressed as 
$S_{cg} < \Lambda^4 R^4$ or
\be
{1\over H^2 R^4}<\Lambda^2 \;.
\label{new1}
\ee
 Applying $H<R^{-1}$, as well as recalling equation (\ref{14}), we obtain
\be
 H<\Lambda < M_p \;.
\label{new2}
\ee
Now --- inasmuch as
$H$, $M_p$ and $\Lambda$ are the only available energy scales ---
one need only
appeal to {\it Occam's Razor} to  see
that equation (\ref{16}) naturally follows.

Taken together, equations (\ref{14}) and (\ref{16})
paint the following picture:  The  ultraviolet-cutoff scale 
is effectively  ``born''  at the Planck time (or, just as suitably, later) 
and then decreases, dynamically, at a rate that is synchronized with
the cosmic  expansion
of the universe.~\footnote{It is perhaps useful to note that, for many relevant
models, $H\sim 10^{-5} M_p$ at the onset of inflation; after
which  $H$ ``rolls''  slowly downwards until the inflationary era terminates 
\cite{Inf}.}  
We will have more to say on this observation below. 

\subsection{Super-Horizon Perturbations}

As inflation proceeds, the quasi-exponential expansion of the
scale factor rapidly drives the perturbations
 outside of the slowly rolling~$^8$ horizon.
(Note that any given perturbative mode has a proper
wavelength that increases linearly with $R$.) 
After exiting, these inflationary perturbations 
are effectively frozen; forming what may be regarded as
a  fluid of classical modes.  Ideally, for current purposes, we
 would like  to estimate the
amount of entropy that has been carried out in this manner at
late (inflationary) times.  But, just like in the preceding case,
  a procedure of
coarse graining is once again called for.
One natural choice is to initially adopt the perspective of a local
(sub-horizon) observer \cite{KVS} and then
trace over the degrees of freedom in the unobservable (exterior) region;
thus leading to what is known as an entropy of {\it entanglement} \cite{Ent}.
One could, in principle, do the analogous  calculation of tracing over the
interior region  
and --- inasmuch as the total state is pure --- 
both subregions would necessarily   
be assigned the  same amount of entropy. 
Hence, any measurement of the entanglement entropy by an interior observer
should  apply equally
well to the frozen super-horizon perturbations.
As it so happens, an explicit calculation of this nature reveals that \cite{ML}
\be
S_{ent}\sim {\Lambda^2\over H^2}\;,
\label{17}
\ee 
where $\Lambda$ is, once again, the
ultraviolet cutoff for the {\it interior} field theory.

Since the super-horizon region dictates that $R>H^{-1}$ ({\it i.e.}, strong
gravity),
we are now in  a regime for which the Hubble entropy bound (\ref{2})
is most appropriate. 
Let us, therefore, consider the ratio
\be
{S_{ent}\over S_{H}}\sim {\Lambda^{2}\over M_{p}^2 H^3 R^3} 
< {\Lambda^{2}\over M_{p}}\;,
\label{18}
\ee
where the inequality follows from $RH>1$. Hence, it
is quite clear that the Hubble bound will always be satisfied; providing
$\Lambda < M_p$, which is now known
to be the case [{\it cf}, equation (\ref{14})].
This outcome can be viewed as further support for the notion that
holography has, as of yet, no {\it obvious} utility  
as a means of   constraining  inflation \cite{FK,Dan,KVS}. 

\subsection{Dark Energetics}

Due to our current state of affairs --- namely, an accelerating
cosmological phase  
with $\omega<-0.75$  ---  there must be some rather 
exotic form of 
negative-pressure matter or  ``dark energy''
that dominates 
the  universe  \cite{DE}.  Let us assume, for the sake of argument,
 that we are  not dealing
with a true cosmological constant.
It then stands to reason 
that  the dark energetics can, just like the inflationary modes, 
be described  by an effective field theory
at sub-horizon (or weakly gravitating)
 scales. And so, by analogy with our previous discussion,
there should also be an ultraviolet cutoff for this effective theory;
namely  [{\it cf}, equation
(\ref{16})], 
\be
\Lambda\sim\sqrt{M_p H_0}\;.
\label{19}
\ee
Here, $H_0$ represents the current day Hubble parameter; roughly
$10^{-60}$ in Planck units. As has been pointed out elsewhere
({\it e.g.}, \cite{CKN}), this line of  reasoning provides a {\it de facto} 
solution
of the ``cosmological constant problem'' \cite{Wei}.~\footnote{The
problem being, simply stated, why is the energy scale
for the dark energy  some $60$ orders of magnitude smaller
than the Planck mass?}
 More to the point, equation (\ref{19}) 
predicts a dark-energy density of
\be
\rho_{\Lambda}\sim\Lambda^4\sim M_p^2 H_0^2\sim 10^{-120}M_p^4\;,
\label{20}
\ee
which does indeed comply with the observational evidence.
Unfortunately, even if there
is more than  an element of truth to this deduction,
it is still conspicuously lacking  an underlying
(dynamical) mechanism.
On the other hand, M. Li 
has suggested that the dark energy could have a purely holographic origin
and, by  taking the infrared cutoff  to be
 the {\it future} event 
horizon,
has  constructed a dynamical framework that is consistent
with the current observations \cite{Li}.~\footnote{Given
this  particular framework, some other logical choices
for the infrared cutoff --- such as the
past event (or particle) horizon or the precise location 
of the Hubble horizon itself ---
turn out to be incompatible
with the observed equation of state \cite{Hsu,Li}.} 
This scheme, although enticing, does however appear rather {\it ad hoc}.
Clearly, further input will be required to motivate this choice. 
(For a preliminary attempt, see \cite{XX}.) 
Alternatively, it could be argued that the field-theory cutoff $\Lambda$
is the ``fundamental''  dynamic entity ({\it cf}, the observation at the end
of Subsection A), which then  acts to fix the infrared cutoff
by way of the equation (\ref{15}). 
If this were the case, one would not
want to identify, {\it a priori}, the infrared cutoff
with any particular horizon; rather, the dynamics of the various horizons
(Hubble, future, particle, {\it etc}.) would best be viewed as
manifestations of a dynamically evolving  $\Lambda$. 
We will continue on with this line of reasoning 
at the very end of the paper.  

\subsection{Final Thoughts}

Having scrutinized three distinct cosmological eras,
we are now in a position to reflect  upon
Myung's proposal \cite{Myu-1} of  an encompassing holographic connection.
Hence, a  few pertinent comments are in order:
First of all, for both inflation and the current accelerating period, there
appears to be a {\it holographic} energy content --- by which we mean
a non-conventional matter source for which the energetics
can be described strictly in terms of an
ultraviolet and infrared cutoff. 
Moreover, it has been argued 
(both here and elsewhere; {\it e.g.}, \cite{CKN}) that the two
cutoff scales will most likely be implicitly related. 
And so, this  holographic picture
really boils down to just  a single cutoff parameter; say, $\Lambda$.
It is then the ({\it holographically induced}) 
dynamics of  $\Lambda$ that would 
enable the proposed connection between the two temporally distant eras.
Another plausible example of a  holographic-inspired connection
follows from an inspection of equation (\ref{17}).
Here, we see that the super-horizon inflationary modes ---
which can be regarded as  {\it classically} frozen --- 
also
 ``know'' about the holographic cutoffs;
this, in spite of the fact that the cutoff parameters are intrinsic
to a {\it quantum} field theory  which is confined to the horizon
 {\it interior}. 
Finally, let us re-emphasize the particular significance of the current work: 
The underlying formalism  has now been  essentially  reduced to just the 
single --- and universally justifiable \cite{Bru-1} ---  holographic bound
of equation (\ref{5}).

Let us finish the paper on a somewhat more speculative note. 
It is, first of all, useful to recall (and then reinterpret) our observation at
 the very end of Subsection A:  
Namely, the cutoff-energy scale appears to progressively decrease
from  the Planck mass 
 --- passing through a value of roughly $10^{-5}$ 
during inflation --- to the present day  value of $10^{-60}$.
Such behavior, a trajectory of ever decreasing scale, is quite reminiscent of 
a renormalization-group
flow \cite{KW}.~\footnote{Note that Strominger \cite{Str} and 
Balasubramanian {\it et al}
\cite{BBM} were the first to  argue, on the basis
of holography, that 
the evolution of the universe could be
interpreted  as such a flow. However, their perspective, not to mention flow
direction, differs from  here.}
To fill out this analogy,  we can regard $\Lambda$  as the
(holographic) $c$-function,
time as the scale factor, and the (perceived)  evolution of the universe
as the  renormalization-group trajectory.
(At least from the viewpoint of a local observer, for whom
$\Lambda$ takes on a definite meaning. A hypothetical global observer
would likely have a much different interpretation, as could also
be anticipated on the grounds of horizon complementarity \cite{BF}.)
This framework has the esoteric appeal of a truly holographic universe,
with one of the spacetime dimensions --- cosmic time ---
 being  an emergent (holographically induced)  construct.
Such an outcome is consistent with the notion that
the fundamental quantum theory of gravity should ultimately have a 
background-independent
meaning \cite{Rov}.
As for the viability of such speculations, presumably
only time (and quantum gravity) will tell.

\section*{Acknowledgments} 

 Research is supported  by
the Marsden Fund (c/o the  New Zealand Royal Society) 
and by the University Research  Fund (c/o Victoria University).
The author remarks that no birds were harmed in the preparation
of this manuscript ({\it cf}, page 4).

\end{document}